# Experiments on Generative AI-Powered Parametric Modeling and BIM for Architectural Design

**Jaechang Ko** [a, *], **John Ajibefun**[a], **Wei Yan** [a, *]

[a] *Department of Architecture, Texas A&M University, College Station, 77840, TX, United States*

**Abstract**

This paper introduces a new architectural design framework that utilizes generative AI tools including ChatGPT and Veras with parametric modeling and Building Information Modeling (BIM) to enhance the design process. The study experiments with the potential of ChatGPT and generative AI in 3D architectural design, extending beyond its use in text and 2D image generation. The proposed framework promotes collaboration between architects and AI, facilitating a quick exploration of design ideas and producing context-sensitive, creative design generation. By integrating ChatGPT for scripting and Veras for generating design ideas with widely used parametric modeling and BIM tools, the framework provides architects with an intuitive and powerful method to convey design intent, leading to more efficient, creative, and collaborative design processes.

1. **Introduction**

The architectural design process is complex and continuously evolving, utilizing various tools and methodologies to support architects in modeling, visualizing, and analyzing their designs (Cross 1982). With the rapid advancement of technology, artificial intelligence (AI) and machine learning (ML) have been integrated into the design process, presenting new opportunities and challenges for architects and designers. However, the potential for AI, particularly language models like ChatGPT - a conversational AI model developed by OpenAI (Radford et al. 2021)- to transform the architectural design process has yet to be fully explored. This paper presents a new framework for architectural design that uses ChatGPT and AI-based ideation and visualization tools, Veras ("VERAS" 2023), to make the design process easier and create 3D geometric models, parametric models, and Building Information Models using natural language input.

The proposed framework combines ChatGPT and Veras to generate and explore design ideas rapidly. Using natural language input, architects can communicate their design intentions more intuitively, allowing quicker iterations and reducing barriers associated with traditional design tools (Hsu, Yang, and Buehler 2022). Moreover, ChatGPT's ability to understand human design intentions helps to translate the input into Building Information Modeling (BIM) and parametric models, highlighting the potential of the architectural design process.

In this paper, ChatGPT extends its capabilities in creating 3D architectural models beyond the 2D images and text generation in traditional methods (Zhou et al. 2023). The paper presents two distinct implementation methods: ChatGPT in BIM with Dynamo, and ChatGPT API within Grasshopper in Rhino. This integration enabled the generation of complex, information-rich, and parametrically controlled 3D architectural models in both frameworks. This allowed us to shift towards more interactive and adaptive design methodologies beyond conventional 3D modeling methods in architecture pedagogies. Subsequently, the two frameworks proposed in ChatGPT are integrated with the Veras, AI-based ideation and visualization tool. This integration allowed us to work more efficiently and effectively in their rapid design exploration by receiving real-time feedback and generating design alternatives. Furthermore, we could engage in a more iterative and dynamic design process by evaluating multiple design options and selecting the best solution to evolve our design.

This paper comprises two main sections. The first section presents a unique implementation method that integrates ChatGPT and parametric BIM with Revit and Dynamo. It outlines their advantages and unique contributions. The second section introduces another implementation that combines ChatGPT API with parametric modeling tools Rhino and Grasshopper. Both workflows leverage Veras, for AI-based ideation and visualization. The paper discusses the potential benefits and compares the experimental results.

## 2.    ChatGPT for BIM Dynamo + Veras

In this section, we explore the implementation of ChatGPT (version 4.0) and Veras in the BIM Dynamo design process. The integration of ChatGPT with Dynamo allows for an intelligent and interactive design process, while the addition of Veras provides valuable feedback to refine the generated design solutions (See figure 1).

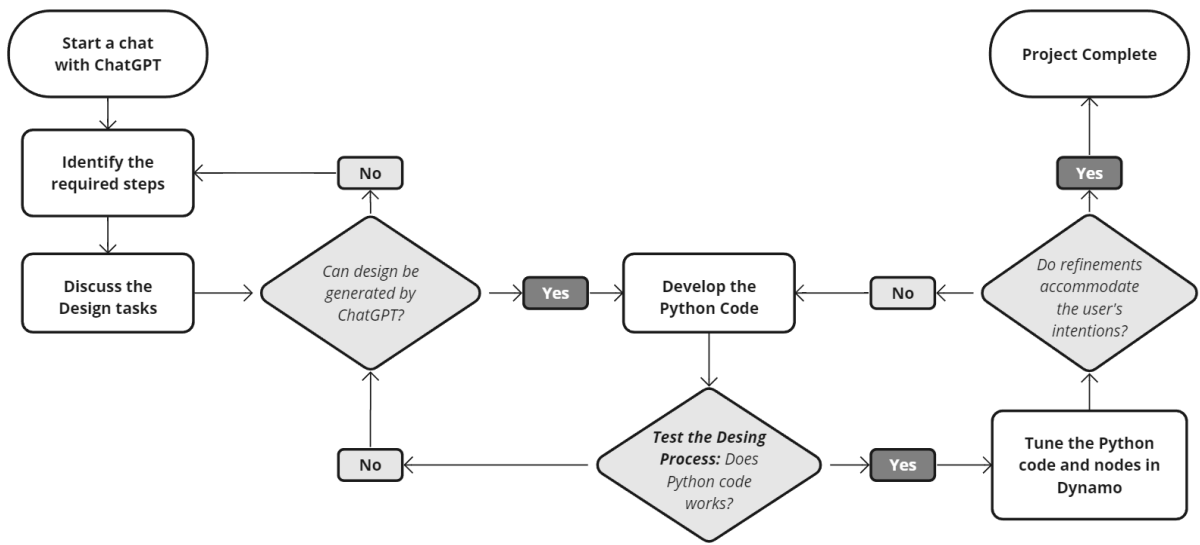

Figure 1. ChatGPT + BIM Dynamo implementation workflow

The process begins with discussing the design task. ChatGPT helps identify the required steps to accomplish the task, leading to the development of Python code as functions in the visual programming environment Dynamo for the BIM authoring tool Revit. The Python code in Dynamo is tested to ensure the design process works as intended. The Python code or process is refined if needed. Once the design results are satisfied with the users, the chat with ChatGPT is terminated.

## 2.1. Step-by-Step Conversations

This section analyzes the interaction between the user and ChatGPT in the design process. The chat history is essential to understand how the AI understands the architect's intentions, converts them into specific instructions, and eventually creates a BIM model of an example building design.

- **Step 1. Initial Query and User Input**

This phase marks the start of the design process, where the user inputs the initial design prompt. The initial prompt could be either a vague architectural concept or a more specific design requirement. We began the interaction with a specific requirement to create a wall with a window in a Dynamo (Figure 2).

```
User
Write python code for Revit Dynamo to create a wall with a window

ChatGPT
Revit Dynamo is a visual programming platform that allows users to create and
modify Revit models through scripts and custom nodes. You can use Python to
create a custom node in Dynamo, which can then be used to create a wall with a
window.

Here's an example of Python code that you can use in Dynamo to create a wall with
a window:

```python
import clr

# Import Revit Nodes
clr.AddReference("RevitNodes")
import Revit
clr.ImportExtensions(Revit.Elements)

# Import Revit Services
clr.AddReference("RevitServices")
import RevitServices
from RevitServices.Persistence import DocumentManager
from RevitServices.Transactions import TransactionManager

# Import Revit API
clr.AddReference("RevitAPI")
import Autodesk
from Autodesk.Revit.DB import *
```

**Figure 2. Initial chat in the experiment (ChatGPT exported data)**



- **Step 2. User Feedback and Troubleshooting**

Verifying and refining the generated Python code is crucial. The authors encountered bugs when running the generated Python code in Dynamo and needed to refine the provided solution. During this phase, the user provides feedback on the code generated by the AI model, which leads to troubleshooting and code refinement. The process is that the user copied the ChatGPT generated Python code into Dynamo, ran the Dynamo program, and normally received error messages. The user then manually typed the error messages back into ChatGPT, which then debugged the code and provided new code (See Figure 3).

```
User
but structuraltype is not defined

ChatGPT
My apologies for the confusion earlier. I made a mistake in my response, and I
appreciate your patience. The import statement was indeed correct, and the issue
is not with the `StructuralType` import.

Here's the updated code with the correct method overload for `NewFamilyInstance`:

```python
import clr
```

Figure 3. Feedback and Troubleshooting

The user then re-copied the new code and replaced the old one in Dynamo, ran the Dynamo program again, and normally the previous bug was resolved, but a new bug appeared. The user then repeated the above process, until all the bugs were resolved (See Figure 4).

```
User
error creating window name structural type is not defined

ChatGPT
My apologies again for the oversight. I missed importing the `StructuralType` class from the
`Autodesk.Revit.DB` namespace. Here's the updated code with the proper import statement:

```python
import clr
```

Figure 4. Debugging Interaction Cycle

Finally the working Python code was completely generated by ChatGPT. Also the debugging is completely done by ChatGPT, except that the user had to copy (type) the error message manually from Dynamo/Python into ChatGPT. The authors expect that this manual work could be automated in the future by integrating ChatGPT into Dynamo through ChatGPT Application Programming Interface (API). Therefore ChatGPT could provide troubleshooting and code modification guidance, making it a versatile tool for interactive problem-solving.



- **Step 3. Chat for Creating BIM**

It started with creating a wall with a window, then progressed to creating a room with four such walls connected, and finally to creating a roof and a door. Each step was built on top of the previous step's outcome. Each step included a Python node with the ChatGPT-generated and debugged code.

This approach allows for building a complex model from simple components step by step. Figure 5 shows the chat for the step of creating the room, after creating a wall with a window.

```
User
This works! Thanks!

ChatGPT
You're welcome! I'm glad the code works for you. If you have any more questions
or need further assistance, please feel free to ask!

User
Can you use the same way to create a room with four connected walls with their
windows (same dimensions)?

ChatGPT
Sure! Here's a Python script to create a rectangular room with four connected
walls and windows on each wall:

```python
import clr
import math

# Import Revit Nodes
clr.AddReference("RevitNodes")
import Revit
clr.ImportExtensions(Revit.Elements)

# Import Revit Services
clr.AddReference("RevitServices")
import RevitServices
from RevitServices.Persistence import DocumentManager
from RevitServices.Transactions import TransactionManager

# Import Revit API
clr.AddReference("RevitAPI")
import Autodesk
from Autodesk.Revit.DB import *
from Autodesk.Revit.DB.Structure import StructuralType
```

**Figure 5. The user asked ChatGPT to create a room, and ChatGPT responded. (Data exported from ChatGPT)**

- **Step 4. ChatGPT Guiding the Design Process**

In this step, ChatGPT takes on a more directive role, going beyond the mere execution of commands. ChatGPT could suggest general modeling processes or even recommend modifications based on its understanding of the chat history (See Figure 6).

We requested ChatGPT to add a new door to the wall provided. ChatGPT utilized code derived from previous prompts rather than creating a completely new process, with modeling steps based on all previous conversation information.

```
User
given a wall, door type, and level, insert a door into a wall, using Revit Dynamo
Python script

ChatGPT
To insert a door into a wall using Revit Dynamo Python script, you can follow
these steps:

1. Start by importing the necessary Revit and Dynamo libraries at the beginning
of your script:

```python
import clr
clr.AddReference('RevitAPI')
from Autodesk.Revit.DB import *
clr.AddReference('RevitServices')
import RevitServices
from RevitServices.Persistence import DocumentManager
from RevitServices.Transactions import TransactionManager
doc = DocumentManager.Instance.CurrentDBDocument
```

2. Define the inputs to the script, which should include the wall, door type, and
level:

```python
wall = IN[0]
doorType = IN[1]
level = IN[2]
```

3. Get the geometry of the wall, and extract the location and orientation
information for the wall:

```python
wallGeometry = wall.GetGeometryObjectFromReference(Reference(wall))
wallLocation = wall.Location.Point
wallNormal = wallOrientation = wallGeometry.Faces.Item[0].Normal
```
```

**Figure 6. ChatGPT suggest modeling process with modifications**

This phase implies an enhanced level of interaction between the user and the AI model, which uses its understanding of design processes and problem-solving strategies to actively guide the user through various stages.

- **Step 5. Interactive Developments**

ChatGPT's capabilities extend beyond fixing errors to include adding new elements or expanding the model. In Figure 7, we requested to add a door and roof to the existing model. ChatGPT provided the updated Python code with four interconnected walls, windows, a door, and a roof additionally.

```
User
Great! This works! Can you add a door and a roof to the building?

ChatGPT
Certainly! Here's the updated code to create a room with four connected walls, windows, a door, and a roof:

```python
import clr
import math

# Import Revit Nodes
clr.AddReference("RevitNodes")
import Revit
clr.ImportExtensions(Revit.Elements)

# Import Revit Services
clr.AddReference("RevitServices")
import RevitServices
from RevitServices.Persistence import DocumentManager
from RevitServices.Transactions import TransactionManager

# Import Revit API
clr.AddReference("RevitAPI")
import Autodesk
from Autodesk.Revit.DB import *
from Autodesk.Revit.DB.Structure import StructuralType

doc = DocumentManager.Instance.CurrentDBDocument
app = DocumentManager.Instance.CurrentUIApplication.Application

# Input from Dynamo
wallType = UnwrapElement(IN[0])
windowType = UnwrapElement(IN[1])
doorType = UnwrapElement(IN[2])
length = IN[3]
width = IN[4]
wallHeight = IN[5]
windowWidth = IN[6]
windowHeight = IN[7]
doorWidth = IN[8]
doorHeight = IN[9]
level = UnwrapElement(IN[10])
```

**Figure 7. Building Expansion and Core Updates**

This demonstrates ChatGPT's flexibility and adaptability to the evolving design needs and goals of the user. Thus, ChatGPT plays a significant role in the iterative and dynamic process of building design, adjusting its responses and code based on user input and feedback.



● **Step 6. Implementation and Fine Tuning**

After generating each code with ChatGPT, users must instantiate it in Dynamo by creating a Python Script node and entering the AI-generated code. Successful implementation of the code requires user engagement in connecting the corresponding input and output nodes in line with the script's requirements. In this experiment of creating the room in Revit/Dynamo, the AI-generated Python code requires input nodes such as wallType, windowType, length, width, wallHeight, windowWidth, windowHeight, and level. Users must determine the specific data type of each input (i.e. a number, building object type, or level) and set the corresponding conditions accordingly (See Figure 8).

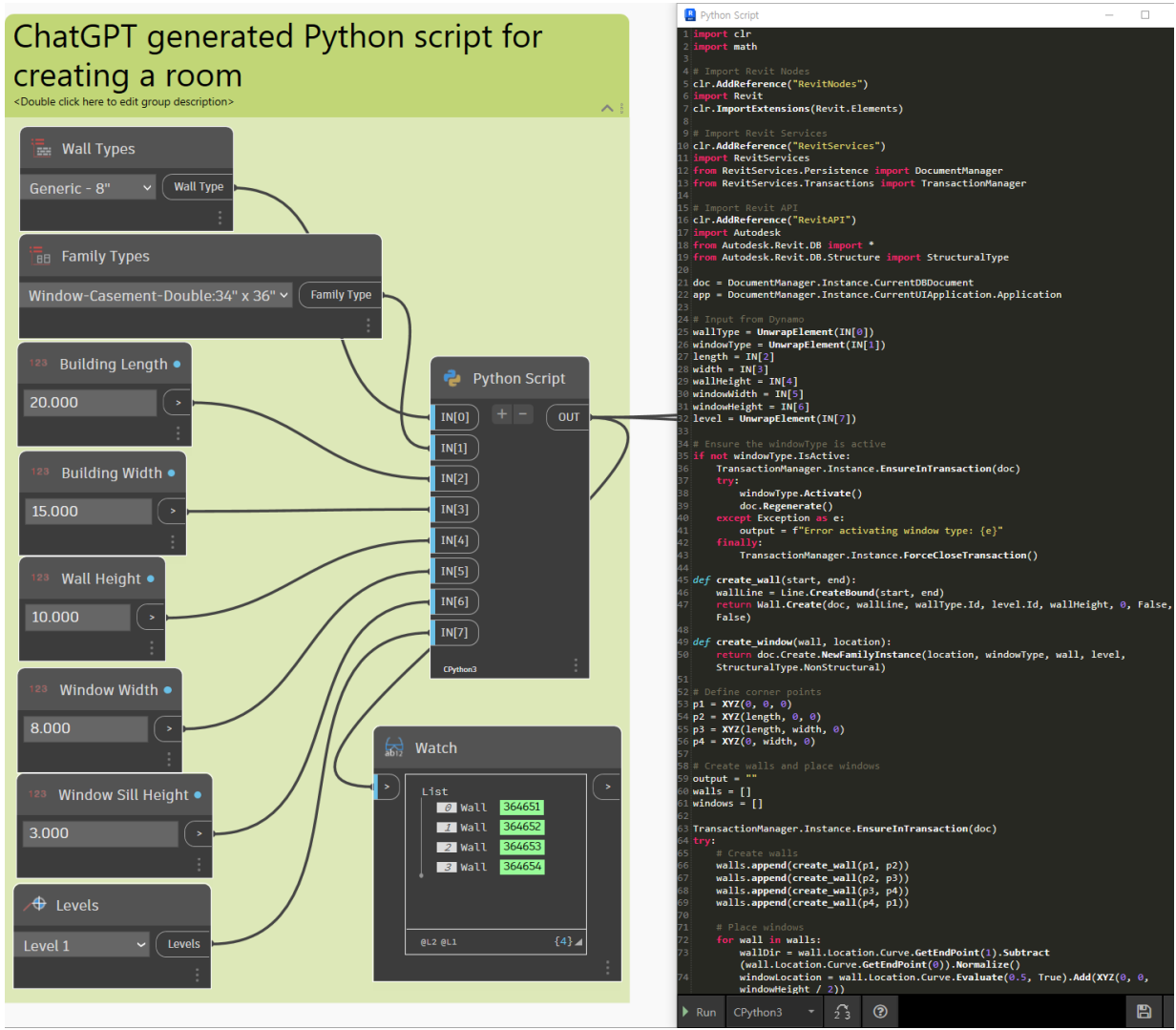

**Figure 8. The Python Script node and other input nodes created by the user manually (left) based on the ChatGPT-generated Python code (right), in Dynamo, which can run to create a building model.**



This demonstrates the potential of ChatGPT in facilitating the architectural design process and shows how AI can enable users to effectively and efficiently explore and realize their design ideas (See Figure 9).

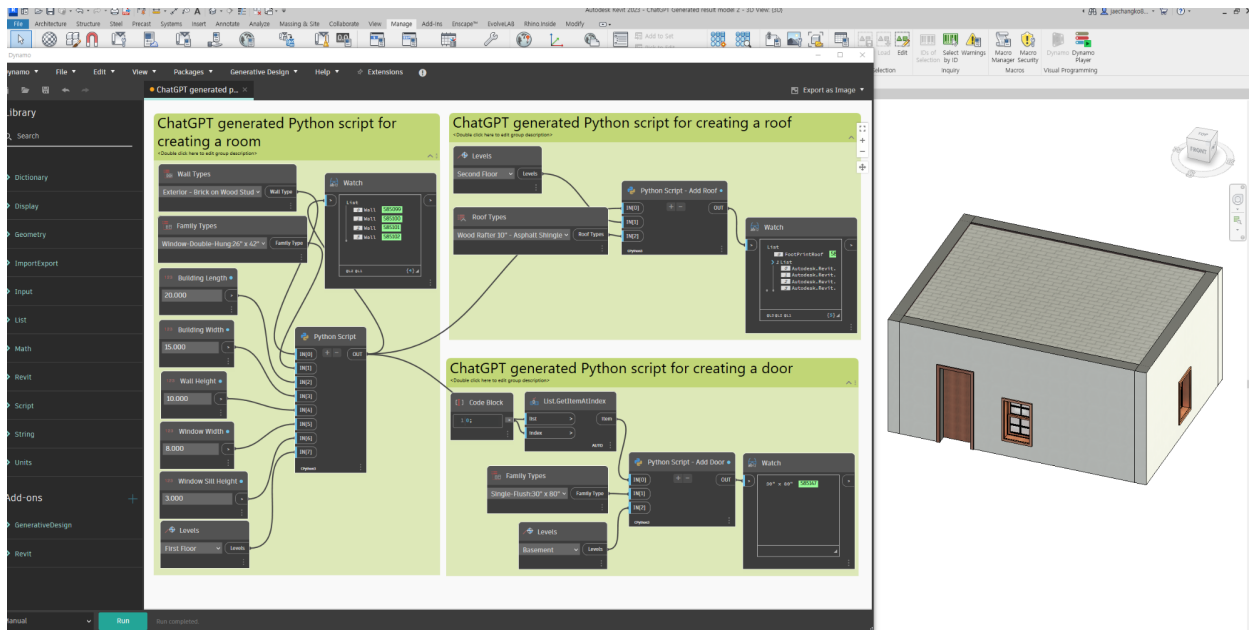

**Figure 9. Final version of Dynamo (left) and ChatGPT-generated BIM (right)**

## 2.2. Integrating Veras for Design Ideation and Visualization

This section discusses the integration of Veras, an AI-powered visualization add-in for SketchUp®, Revit® & Rhinoceros®, that uses the 2D views of modeled 3D geometry as a substrate for creativity and inspiration, into the ChatGPT + BIM process to facilitate the design process. The initial model was a simple box-shaped space featuring four walls, a window on each wall, a roof, and a door. While this serves as a useful starting point, initiating a new design task from this simple model, particularly with limited design experience or unfamiliar tools, can be arduous and exhausting. A significant challenge lies in the explicit semantics of design elements such as walls, windows, doors, roofs, etc., as required by Revit for BIM. This requires a basic understanding of architectural principles, posing a significant barrier for students and designers new to these platforms.

Veras facilitated the generation of diverse design visualizations via user prompts, simply by specifying the material and purpose of the building. For instance, Rendering A in Table 7, prompted with *"Wooden retail store"*, yielded a unique wood pattern direction for the walls and doors and suggested fluctuation in wall sections, providing a information for subsequent design and modification whereas Rendering B, prompted with *"Concrete Office Building"*, reflected dark glazing and setback openings.

This feedback loop extends beyond exterior design to interior spaces. For instance, Rendering C & D(for a different BIM model of the authors), prompted with "*Interior contemporary living room with white paint and wood as major material*", proposed filling the sunken living space and



replacing the TV wall with an open design, and removing the skylight windows for a flat finish (Table 1) or keeping the skylight, depending on the material applied to the ceiling surface.

Table 1. Based on the views from BIM (left column), and the user's prompts, Veras generated different designs' visualization (middle and right columns)

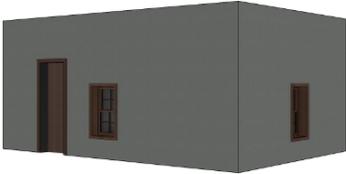

## 3. ChatGPT API in Rhino Grasshopper + Veras

This section explores the integration of the ChatGPT API with Rhino Grasshopper to generate 3D architectural models using natural language input. We discuss integrating the ChatGPT API with Grasshopper and generating prompts as components.

### 3.1. Import ChatGPT API in Rhino Grasshopper

We incorporated a trigger component and data recorder in Grasshopper to create an iterative design process. The trigger component facilitated the automatic and continuous execution of the ChatGPT API requests. By setting the trigger interval (e.g., 2 seconds), we ensured the API request would prompt every interval, allowing us to generate various designs rapidly.

While the API requests were created in every interval, the Grasshopper script also constructed new coordinates per iteration, which allowed us to generate new geometry with each trigger.



## 3.2. API as Components

Our objective was to create actual geometries without text, human manual copy and paste, or additional manual cleanup processes. We used 'loft' at the end of the process to generate the final geometry. Initially, we attempted to create polynomial curves using the prompt: *"Create one polynomial with three variables x, y, and z."* We observed that the generated equations included numbering and text, which was undesirable for our purposes since we needed only equations (See Table 2). Therefore, we added two additional prompts: *"Do not number the equations"* and *"No text, only equations."*

Table 2. Prompt developments process

| | Initial Prompt | Prompt Addition 1 | Prompt Addition 2 | Prompt Addition 3 |
|---|---|---|---|---|
| Text | *"Create one polynomial with three variables x, y, and z."* | *"Do not number the equations"* *"No text, only equations."* | *"Use the * operator whenever multiplication occurs."* | *"create one trigonometric polynomials with three variables each x, y, and z"* *"Only use sin and cos function not tan."* |
| Panel Results | {0;0} 0. xÂ3 + 2xyz + 5yÂ2z - 7zÂ3 | x^2y + 2xyz + z^3 | x*y*z + 2*x*y + 3*x*z + 4*y*z + 5*x + 6*y + 7*z + 8 | sin(x)*cos(y)*cos(z) + cos(x)*sin(y)*sin(z) |
| Output | N/A | N/A | | |

To create more consistent, regular, and patterned curves, we introduced constraints to generate trigonometric polynomials using only sine and cosine functions. This approach led to the successful generation of polynomial curves, avoiding drastic irregular geometries. To further examine ChatGPT's understanding of shape descriptions, we tested again based on two different prompts: 1) *"Generate a polynomial curve that has placid, calm, and linear waves."* and 2) *"Generate a polynomial curve that has surge, drastic, and crazy fluctuation waves"*.

Figure 10 shows the results regarding ChatGPT API generating two distinct shape designs easily recognizable by humans.

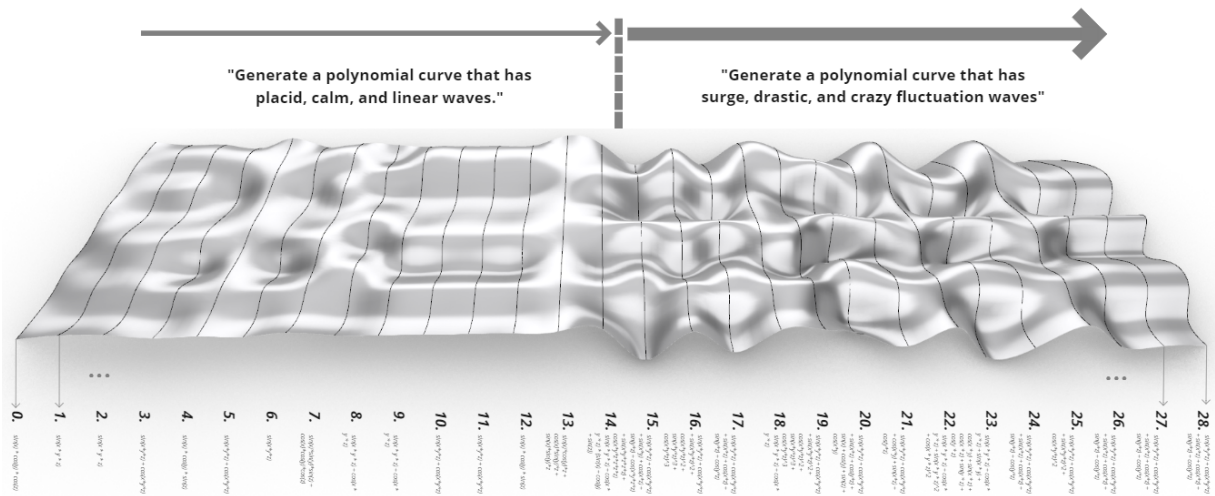

Figure 10. Descriptive language implementation in ChatGPT API



This outcome demonstrates that ChatGPT can interpret human language input and generate designs that cater to specific requirements or emotions. It highlights the importance of prompt for unlocking the full potential of ChatGPT-generated architectural design, ultimately expanding the possibilities in design methods.

### 3.3. ChatGPT as Generative Design Tool

We further investigated ChatGPT's ability to adapt to regulations or constraints in the design process. We aimed to create closed curves that could be lofted together to form linear buildings. We prompted the API to generate 10-sets of coordinates for a curve in the xz plane, keeping the y-values constant at 0 to maintain alignment. However, this approach led to some twisted curves, which resulted in conflicts when lofting the shapes together (See column a in Table 3).

Table 3. Prompt with topological condition

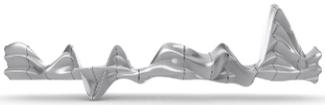

Therefore, we revised the prompt to focus exclusively on generating curves, including phrases such as *'ensuring convex'* and *'avoiding intersecting lines'*. This refinement allowed ChatGPT to produce a range of dynamic, non-intersecting, and convex (or concave, depending on the viewing direction) shapes. The API adapted to these new constraints quickly and effectively, highlighting ChatGPT API's potential in creating diverse and complex geometric forms, adhering to specific design constraints.

Then, we introduced additional parameters for column generation. We stipulated that the generated curve must always incorporate a circle with a radius of 6 units, serving as a structural minimum cross-section framing to ensure stability. Simultaneously, the center point of the interpolated convex curves could vary, but it had to be contained within a circle of radius 3, centered at the origin. This introduced a randomized but controlled element in the design process, maintaining structural constraints. This process shows its versatility and potential in creating designs that satisfy practical architectural requirements.



## 3.4. Veras Visualization

Integrating the Veras allowed for rapid exploration of architectural style, locality, scale, materiality, and context.

Initially, a building was rendered in Veras without specifying any user prompt. This resulted in a building with an aluminum skin set against a broad, flat terrain with a sunset background (see rendering A in Table 4). Intriguingly, Veras autonomously introduced a column to the lower portion of the lifted mass, providing structural supplements.

**Table 4. Veras renderings for design exploration**

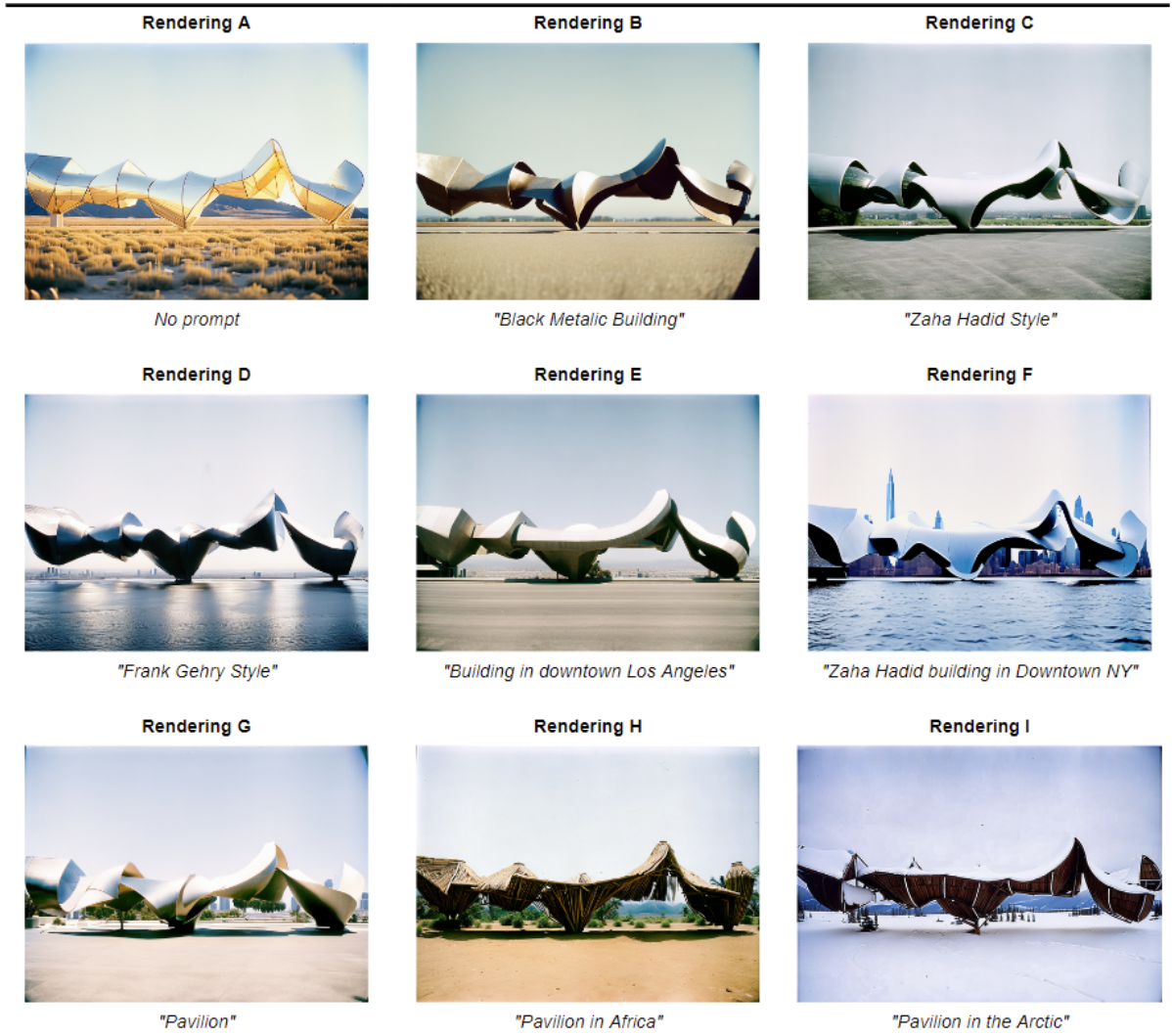

Subsequently, we specified '*Black metallic building*' as a user prompt to evoke a darker, heavier architectural aesthetic (rendering B). Veras responded by creating a building with a more abstract, thicker skin, with the edges of the building clearly defined. An entry was introduced at the lower center of the building, the only section in contact with the ground, further contributing to its integrity.



Inspired by the works of renowned architects like Zaha Hadid and Frank Gehry, we prompted Veras to create buildings in their styles (rendering C and D). The results presented the ability to emulate distinct architectural styles, generating designs of famous architects' iconic works.

Veras's adaptability was tested by inserting locality-specific keywords into the prompt. For instance, based on the keyword '*Building in downtown Los Angeles*', Veras created a background of LA for a concrete and metal-skinned building (rendering E). Similarly, a prompt of '*Zaha Hadid building in Downtown NY*' resulted in a Manhattan-like background with a cool-toned image rendering (rendering F).

Scale features were also explored (rendering G). The keyword '*pavilion*' shrunk the scale by changing the entourages' scale and focused on the building's external skin and structural details, such as stair-like features in the openings and structural elements supporting the lifted mass.

Finally, regional and material-specific factors were assessed. The same 'Pavilion' material was applied to different regional contexts, such as Africa and the Arctic (rendering H and I), and each rendering responded distinctively to its context. When a pavilion was rendered with the prompt of 'Africa', it suggested a bamboo structure with wide spacing for ventilation and airflow. On the other hand, in the Arctic, the bamboo spacing was rendered densely to imply a more compact structure, with snow accumulated on its surface.

### 3.5. Veras Design Feedback

The ability of Veras to provide immediate visual feedback proved to be an effective tool in the iterative design process. It was particularly useful in refining and realizing more complex geometries generated through the ChatGPT API (See Table 5).

**Table 5. Comparative Visualization of Design Iterations: basic prompt(left), addition1(middle), and addition2(right)**

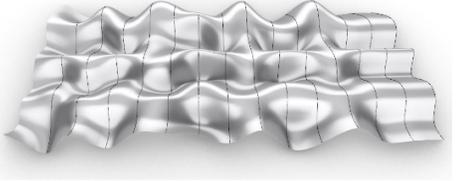
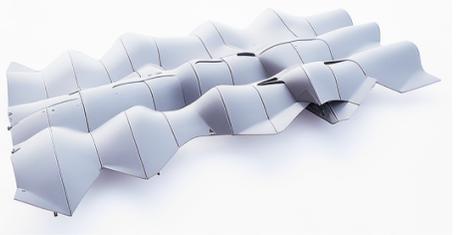
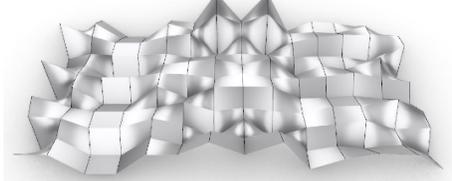
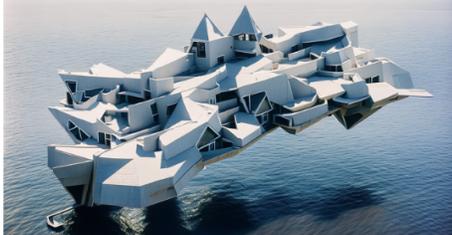



This was observed when rendering the facade created by the ChatGPT API using the prompt *"Generate a polynomial curve that has surge, drastic, and crazy fluctuation waves"* (See middle column in Table 5). Veras translated this wave-shaped design into a rendering as a shell-shaped roof. Although this result was intriguing, we wanted to create a more angular and protruding poly shape output, which prompted a design revision.

We modified the instruction to the ChatGPT API by adding *'interpolated curve degree is 0'*.(See right column in Table 5). The intention was to emphasize sharp angles and discontinuities rather than smooth curves. The API's response to this modified prompt was a markedly different model featuring an angular, poly-shaped structure. The image produced portrayed a more fragmented and jagged architecture, distinct from the initial shell-shaped roof design. This process highlights the power of AI-based Visualization tools for rapid visual feedback and iterative design. Veras effectively facilitated visualization in the design process by generating immediate designs created by the ChatGPT, enabling designers to iterate and refine their ideas rapidly.

## 4. Results and Discussions

In the conventional design process, architects and designers have to imagine and model the elements of the building model, such as walls, ceilings, roofs, and site models. They then need to create renderings to visualize the design, get feedback, and refine the model, repeating this iterative process until a satisfactory design solution is fulfilled. This repetitive process is not only labor-intensive but also time-consuming. This can lead to increased project complexity and budget, requiring constant collaboration and real-time updates from all team members. Furthermore, this iterative modification significantly escalates the potential errors, posing additional management, computational challenges, and complexity.

The novel design process proposed in this study integrates ChatGPT, its API, BIM, parametric modeling, and Veras, resulting in a more efficient and creative workflow. This integration reduced the labor and time typically consumed in the iterative design process for coding in Dynamo. Furthermore, the ChatGPT API in Grasshopper generated codes and polynomial equations based on descriptive language prompts, facilitating the creation of complex and diverse geometric shapes. Subsequently, implementing Veras offered immediate and realistic feedback on designs, enabling a quicker and more effective iteration process. The AI-based rendering tool's ability to manipulate architectural style, scale, materiality, locality, and context greatly enhanced the design exploration process. This significant shift in the design approach reduced the time and complexity of creating and adjusting Grasshopper scripts, enabling designers to focus more on creative exploration and less on technical scripting.

Our exploration of ChatGPT and Veras in architectural design applications reveals promising potential but also uncovers critical challenges for further research. Specifically, it is still unclear how well the current generative AI can address complex, bespoke, or customized space design Since these design intricacies are crucial to generative design  (Nagy et al. 2017), a more comprehensive study into how AI can effectively accommodate them is imperative. Furthermore, we cannot overlook the physical feasibility and structural integrity of AI-generated designs. The Veras implementation process highlighted some cases where AI-generated designs resembled



illusionary constructs like Penrose's Stairs (Nakatsu, Takahashi, and Moriya 2013). Although visually impressive, these designs were architecturally infeasible, reminding us that the boundary between perception and reality in AI-guided design requires further exploration.

## 5. Conclusions

The journey towards realizing the full potential of AI in architectural design is ongoing. This study has demonstrated that integrating technologies like ChatGPT and Veras into the architectural design process can enhance efficiency, collaboration, and creativity. However, this new frontier in architectural design presents challenges that require continual exploration and refinement. As AI technologies continue to evolve, their capabilities for generating unique and novel design solutions will also increase, pushing the boundaries of what is conventionally considered feasible or imaginable in architectural design. Despite the immense potential, it is crucial to recognize the challenges associated with these technologies. The limitations identified in this study should serve as catalysts for future research and development efforts. As we navigate the evolving landscape of AI in architectural design, a continued scholarly and industry-wide dialogue is critical. This dialogue, guided by rigorous empirical research and inclusive professional discourse, will be instrumental in fully harnessing the potential of AI in architectural design.

## 6. Future work

There are many potential research directions based on our study. For example, while Veras can generate many new design ideas and their visualizations as 2D images based on the Revit or Rhino models' 2D views, future research project will investigate the reversed process: how to create 3D parametric BIM based on the new design ideas generated by Veras or other AI tools? As a final experiment for this paper, we thought: why not ask AI to help us look into the future? When we asked ChatGPT on May 24, 2023: *"What is the future of generative AI for architecture?",* here's what ChatGPT responded (Figure 11):



What is the future of generarive AI for architecture?

The future of generative AI for architecture holds great potential and is expected to have a significant impact on the field. Here are some potential developments and trends that may shape the future:

1. Design Assistance: Generative AI can assist architects by providing automated design suggestions and generating multiple design alternatives based on predefined criteria. It can analyze large datasets of architectural designs, understand patterns, and propose novel solutions, helping architects streamline their design process and explore more creative possibilities.
2. Optimized Designs: Generative AI can help optimize architectural designs for factors such as energy efficiency, structural integrity, and material usage. By simulating and analyzing various design configurations, AI algorithms can identify optimal solutions that balance functionality, sustainability, and aesthetics, leading to more efficient and environmentally friendly buildings.
3. Site-Specific Designs: AI algorithms can consider site-specific parameters, such as location, climate, terrain, and cultural context, to generate designs that are tailored to the specific requirements of a particular site. This approach allows architects to create buildings that harmonize with their surroundings and respond to local conditions effectively.
4. Human-AI Collaboration: Generative AI tools will facilitate collaboration between architects and AI systems. Architects can leverage AI algorithms as creative partners, using them to augment their own design expertise and generate ideas. The synergy between human creativity and AI computational power can lead to more innovative and efficient architectural solutions.
5. Virtual Reality and Augmented Reality: Generative AI can integrate with virtual reality (VR) and augmented reality (AR) technologies to provide immersive experiences for architects and clients. Architects can visualize and interact with generated designs in real-time, enabling them to better understand spatial relationships, test design concepts, and communicate ideas effectively.
6. Sustainable Design: As sustainability becomes increasingly important in architecture, generative AI can contribute to the development of eco-friendly designs. AI algorithms can optimize building layouts, materials, and systems to minimize energy consumption, maximize natural lighting, improve indoor air quality, and incorporate renewable energy sources, promoting sustainable and green architecture.
7. Adaptive Architecture: Generative AI can enable the creation of adaptive buildings that respond to changing user needs and environmental conditions. By analyzing user behavior, AI algorithms can generate designs that can adapt over time, accommodating different functions, spatial configurations, and occupancy patterns, leading to more flexible and responsive architectural solutions.

It's important to note that while generative AI offers exciting possibilities, human expertise and creativity remain crucial in the architectural design process. The future of generative AI in architecture is likely to involve a harmonious collaboration between architects and AI systems, leveraging the strengths of both to create inspiring and sustainable built environments.

**Figure 11. GhatGPT Response about the Future of Generative AI for Architecture**



We think that ChatGPT responded well to the major future research directions. Currently researchers in related fields could tackle the challenges in these directions. The authors are working in some of the directions, including optimized design, human-AI collaboration, Virtual Reality and Augmented Reality, etc.